# Forecasting the Integration of Immigrants

Pierluigi Contucci[1], Rickard Sandell[2], and Seyedalireza Seyedi[1]

[1]Dipartimento di Matematica, Università di Bologna,

[2]Departamento de Ciencias Sociales, Universidad de Carlos III de Madrid.

## Abstract

This paper presents a quantitative framework for forecasting immigrant integration using immigrant density as the single driver. By comparing forecasted integration estimates based on data collected up to specific periods in time, with observed integration quantities beyond the specified period, we show that: Our forecasts are prompt—readily available after a short period of time, accurate—with a small relative error—and finally robust—able to predict integration correctly for several years to come. The research reported here proves that the proposed model of integration and its forecast framework are simple and effective tools to reduce uncertainties about how integration phenomena emerge and how they are likely to develop in response to increased migration levels in the future.

**Introduction**

The research reported here addresses the problem of forecasting immigrant integration. The question we are concerned with is; if immigration rise by say 2 to 3 % would this affect the level of immigrant integration in the social system under study? If so, what is the effect magnitude? And finally, how precisely can we forecast it? To this end we develop and introduce a theoretical and mathematical forecasting framework. We expose the resulting model to exceptionally rich integration data from Spain. The empirical tests show that our model deliver precise forecasts of future immigrant integration levels, in multiple integration contexts, and for the entire timespan for which data is available.

Uncovering ways to forecast integration is a potentially important task. The integration of immigrants is and has long been a political priority in countries receiving immigration (European Commission 2005; Penninx et al. 2008; Jacoby et al. 2013). For example, achieving a minimum of integration is considered a necessity, to avoid friction and conflict between immigrants and natives in the host society (Castles and Miller 2009; Niessen and Huddleston 2010; European Commission 2011; IOM 2011; European Commission 2014). Furthermore, and perhaps most importantly, as the effects of low birth rates and aging populations are becoming manifest, high levels of immigrant integration or assimilation is considered by some of the world's leading economies key in building a competitive and sustainable economy for the future (European Commission 2010; European Council 2010; Giovagnoli 2011; Canada Government 2012).

However, the capacity to formulate effective integration policies hinges on the availability of scientific theories and works generating strong predictions of how the integration phenomenon is likely to unfold with the passing of time, and in the face of changing levels of immigration. While there is a rich demographic, sociological and economic literature on individual integration outcomes (see, for example, Van Tubergen (2006) for a good overview of this literature), the problem of forecasting the level of integrations in a society is largely ignored by past and contemporary research. Hence, by filling an important gap in the literature on immigrant integration, the research reported here constitutes a significant contribution to our knowledge about integration and integration phenomena.

The paper is organized as follows. First we introduce the integration concept we aim at forecasting. Thereafter we introduce the theoretical and mathematical model of the integration phenomena on which we base our forecasting method, followed by data description. Then we present the methodological procedure, produce the forecasts, and conduct the proper evaluation of these forecasts. Finally, we conclude with a brief discussion of the wider implications of our results.

**Immigrant integration**

What is immigrant integration and what is it that we set out to forecast? Instead of engaging in the complex task of defining and operationalizing integration, we feel comfortable to adopt a pragmatic approach. For example, there is a consensus within and across demographic, sociological and economic disciplines that inter–marriage/inter–partnerships and immigrant labor force participation are core measures of social and socio–economic immigrant integration (see Gordon 1964; Alba and Nee 1997; Raijman and Tienda 1999; Van Tubergen 2006 for a more extensive discussions on the relevance of these integration quantifiers). Therefore, in the research reported here we set out to develop a quantitative forecasting model capable of predicting integration in the form of inter–partnerships and immigrants labor market participation. More specifically, we have suitable data available for the following four integration quantifiers; 1) rate of inter–marriage between native and foreign born, 2) rate of newborns with one native and one foreign parent, 3) rate of indefinite labor contracts given to immigrants, and 4) rate of temporary labor contracts given to immigrants (see further in the section Data Description below). The former are example of social integration quantifiers, while the latter are examples of so called socio–economic integration quantifiers.

**A simple model of a complex phenomenon**

Forecasting integration requires a reliable and efficient model of the integration phenomena. Past research has shown that both individual and contextual factors are important when explaining variation in immigrant integration outcomes (Van Tubergen and Maas 2007). For example, systematic differences in individual integration outcomes are often associated with age, sex, education, language proficiency, and other individual characteristics (Stevens and Swicegood 1987; Hwang et al. 1997; Cortina et al. 2008; Qian et al. 2012). Likewise, contextual factors, such as culture, ethnic–group–size, and local heterogeneity levels are influential when assessing inter–marriage propensities and labor market participation across different groups (Blau et al. 1982; Blau et al. 1984; Portes and Zhou 1993; Qian et al. 2012).

However, here our aim is not to forecast individual integration outcomes. Rather, we are interested in forecasting the trend in integration. That is, we are foremost concerned with the systemic evolution of integration quantifiers, in the presence of "noise" such as factors unique to individuals and their context.

Forecasting the trend in integration calls for a different type of model than what we typically see in contemporary research articles on integration and assimilation. In recent years, there have been considerable efforts to apply ideas and techniques from statistical physics to other areas of science such as economic, finance, social science, and biology. Explaining the social phase transition, modeling collective animal behavior, like natural flocks of birds, and predicting trends and crisis in the financial Markets are some well–known examples of this venture (see Mantegna and Stanley 1999; Castellano et al. 2000; Bouchaud and Potters 2004; Levy 2005; Ballerini et al. 2008; Contucci et al. 2008; Stanley 2008; Bialek et al. 2012). This genre of models has recently been used to study also migration and integration phenomenon (Contucci and Giardina 2008; Barra and Contucci 2010; Barra et al. 2014).

Barra et al. (2014) have produced a theoretical framework of an immigrant–native system in which they successfully model the systemic integration process for social and socio–economic integration quantifiers as a function of immigrant density ($\gamma$).[1] In this work, and using integration data from Spain, they showed that a complicating, but yet intriguing, factor when modelling integration as a function of $\gamma$, is that while labor participation rates grow proportionally to immigrant density, the rates of inter–partnership have a growth law proportional to the square root of the immigrant density.

The different growth processes have well accepted individual solutions. For example, in a two–group system such as a society composed of immigrants and natives, there can be in–group or cross–group couplings. In other words the choice between, say, marrying or hiring an immigrant over a native is dichotomous. A natural candidate to describe the frequency of cross–group couplings in large populations is McFadden's Discrete Choice theory (McFadden 2001). A crucial assumption in the discrete choice theory is mutual independence between the involved random variables (Gallo et al. 2009). Consequently, McFadden's theory would predict linear growth in integration over immigrant density— $\gamma$. And indeed, McFadden's theory works very well when assessing the level of labor market integration as a function of immigration density (Barra et al. 2014). Hence, based on this finding it can be argued that the decisions of contracting an immigrant are made in a mutually independent fashion, regardless of how other actors have decided in this matter[2].

---

[1] It should be noted that the causal relationship between immigrant group size and integration has received ample attention in Blau et al. 1982 and Blau et al. 1984. Blau et al's findings are not contradictory to Barra et al. (2014) work, but complementary. However, in difference with Blau et al, Barra et al model the interaction component explicitly.

[2] Note that the absence of interaction in labor market participation couplings does not mean that social networks are irrelevant in the job market. What the data analysis suggests is simply that the propensity of hiring immigrants over natives is unaffected by whether other employers have or have not hired immigrants over natives before. Social

However, the square root growth in the rate of inter–partnerships suggests that the choice of partner is not well described by the classical discrete choice theory. A plausible source for this mismatch is that in difference to decisions in the labor market, inter–partnership decisions are not taken independently (Weber 1978; Barra et al. 2014). That is, the action of marrying an immigrant or having a child with an immigrant is a decision that is contingent of how others in the environment have acted or not acted in this context before the decision at hand (see also, for example, Kalmijn's (1998) discussion of social determinants in inter–marriage). Theories that relax the assumption of independence and cater to this type of social action and interaction, introduced by Brock and Durlauf (2001) and thereafter further developed (Contucci et al. 2008; Contucci and Giardina 2008; Gallo et al. 2009), predict a square root behavior of the probability of cross–group couplings. This finding is consistent with the growth law observed for the inter–partnership data by Barra et al (2014). That is, data suggest that social network activity induces the intensity of inter–partnerships.

The novelty introduced by Barra et al. (2014), is a mathematical framework that makes it possible to model the two different growth processes in a unified manner. The proposed solution is a generalization of the so called monomer–dimer model (Heilmann and Lieb 1970) with the addition of an imitative interacting social network component with random topology in agreement with the small world–scenario (Watts and Strogatz 1998). The resulting model reduces to the classical discrete choice theory with linear growth in situations when social action and interaction in the integration process is negligible, and to the square root behavior when such factors are prevailing (Barra et al, 2014). When confronted with empirical data the model provides an extraordinary fit across all integration quantifiers regardless of the a priori differences in the importance of social action and interaction discussed above[3].

---

networks are still likely to be determinant for the individuals' chances of landing a job (see Granovetter 1974 on the latter).

[3] See Barra et al. (2014) for a full account of the model, proofs, and empirical tests.

Consequently, it can be shown that the level of integration in a society can be modeled mathematically as a function of immigrant density— $\gamma$ (Barra et al. 2014). Moreover, if the goal is to obtain estimates of the development of the general level of integration in a society as a function of $\gamma$ the model proposed by Barra et al. (2014) does a better job than the currently available alternatives. In what follows we will draw on this work as we develop a theoretical and quantitative framework to forecast integration as a function of immigrant density— $\gamma$.

## Data description

We use immigration data from Spain on the time interval 1999 to 2010[4]. Data on local immigrant densities are compiled as follows. We use the size of the immigrant population and the native population in each municipality as reported in the 2001 Census as our baseline. We then estimate the local immigrant densities for different points in time (quarterly) based on the information contained in the Statistics over residential variation in Spanish municipalities, the so called Estadistica de variaciones residenciales (EVR) as elaborated by Spain's National Statistical Agency (INE). An exceptional feature of the Spanish data is that they also include so–called undocumented immigrants, that is, immigrants who lack a residence permit (see Sandell 2012).

---

[4] This period corresponds to the period in which Spain received most of its current immigrant population For those unfamiliar with the Spanish immigration context, the following brief information may be useful. In 1999, Spain received fewer than 50,000 new documented and undocumented immigrants. Since then, annual immigration levels have increased dramatically, reaching a peak in 2006 and 2007, with inflows exceeding 800,000. Spain's documented and undocumented foreign born population has risen from little more than 1 million to over 6.5 million in the analyzed period. Its share of the total population has risen from less than 3% to over 13% in the same period. Currently there are immigrants from almost all nations in Spain. However, some 20 immigrant origins account for approximately 80% of Spain's total immigrant population. Immigrants from Romania form the largest minority in Spain (767,000 at the end of 2008), followed by immigrants from Morocco (737,000 at the end of 2008) and Ecuador (479,000 at the end of 2008). Europe and South America together account for over 70% of Spain's total immigrant population.

Data on marriages and births are drawn from the local offices of Vital Records and Statistics across Spain (Registro Civil), and have been compounded by the INE. By our definition, inter–marriage occurs when a Spanish–born (native) person marries a person born in a foreign country. Similarly, we consider all births with one native and one foreign born parent to be mixed births. Inter–marriages and mixed births where the foreign–born spouse or parent is an undocumented migrant is included. We focus our analysis on birth and marriage events that occurred during the period 1999 to 2008. However, data on density, marriages and births are subject to data protection restrictions. An individual residence municipality is only disclosed if its population is larger than 10,000. For this reason, out of approximately 8,000 municipalities in Spain, our analysis focuses on only 735. Still, some 85% of Spanish immigrants reside in the included municipalities.

Data on labor contracts come from Spain's Continuous Sample of Employment Histories (the so called Muestra Continua de Vidas Laborales or MCVL). It is an administrative data set with longitudinal information for a 4% non–stratified random sample of the population who are affiliated with Spain's Social Security. Sampling is conducted on a yearly basis. We use data from the waves 2005 to 2010. The inclusion of an individual in the sample is determined by a sequence in the individual's social security number that does not vary across sample waves. This means that individuals are maintained across samples. New affiliates with a social security number matching the predetermined sequence are added in each new wave. The data contain information on contractual conditions such as whether the individuals have a temporary or indefinite labor contract, as well as the contracts start and stop times. Residential data at the level of municipality and information about place of birth are also available. In contrast to the data on densities, marriages, and births, for these data the residence municipality is only disclosed if the population is larger than 40,000.

Operatively, we derive two datasets based on the information described above. One contains data on inter–marriages and mixed births, and the other on labor market participation. Both datasets contain spatial and temporal information, such as the municipal code, quarter, year, and the immigration density in the municipalities across time. The data on labor contracts consist of 3,553 entries over the period 2005–2010. The data on marriages and newborns consist of 27,144 (municipality/time) entries spanning the period 1999–2008.

In our study, we define the integration quantifiers as a function of immigrant density, $\gamma = \frac{number\ of\ immigrants}{total\ population} \in [0,1]$, in the form:

$$M_m = \frac{the\ numbers\ of\ mixed\ marriages}{the\ numbers\ of\ marriages} \tag{1}$$

$$B_m = \frac{the\ numbers\ of\ newborns\ with\ mixed\ parents}{the\ numbers\ of\ newborns} \tag{2}$$

$$J_p = \frac{the\ numbers\ of\ permanent\ contracts\ given\ to\ immigrants}{the\ numbers\ of\ permanent\ contracts} \tag{3}$$

$$J_t = \frac{the\ numbers\ of\ temporary\ contracts\ given\ to\ immigrants}{the\ numbers\ of\ temporary\ contracts} \tag{4}$$

**Method and results**

To prepare the data for analysis, we use the following algorithm: we began by aggregating and cleansing the data in which the records should be checked for missing or incomplete data to avoid inaccuracy in our forecasting approach. After that, data on our quantifiers are organized into two datasets. The first contains information on our social integration quantifiers (see equation 1 and 2). The second set contains data on socio–economic quantifiers (see equation 3 and 4). Next, for each data–set, we perform a detailed test of the methodology's forecast performance for increasing intervals of time: period 1 = (2000 to 2001), period 2 = (2000 to 2002), period 3 = (2000 to 2003), etc. The objective is to forecast; 1) the growth law determining the integration process (i.e., establishing whether it is individual independent action or social action that drives the integration process) and 2) the level of integration in the system under study at different time intervals. More precisely, we set out to evaluate the quality of the forecasts in terms of *promptness*—how long is the waiting time before we can correctly identify the growth law for each integration process— *accuracy*—how well the prediction replicates the observed integration value—and finally *robustness*—the forecasting ability of our model over the entire time span.

Starting from data, we extract the growth law of each quantifier in terms of $\Gamma = \gamma(1 - \gamma)$[5]. Hereupon, we create a data set for each quantifier by merging all the values of $\Gamma$ with their corresponding integration quantifier values. We then order the data by increasing values of $\Gamma$, regardless of their corresponding space and time coordinates. We proceed by grouping the data into bins in terms of $\Gamma$, [6] and finally we compute the averages[7]. In order to avoid noisy result, the number of bins is decided after detailed test of different width of bins. The results of these tests reveal that 5 to 15 bins optimize 581–3471 entries on the socio–economic quantifier's data–groups in defined successive intervals, whereas 8 to 30 bins optimize 2421–26546 entries for the datasets on social quantifiers in determined consecutive periods. Moreover, each bin on socio–economic quantifier represents the average values of 65–248 data versus 220–982 data for social indicators as time progresses (for more details see the tables 1 and 2).

---

[5] The control parameter $\Gamma$ tunes the total number of possible cross-link couplings between immigrant and native populations (see Barra et al. 2014).

[6] For the binning criteria, we used and tested the *constant information* approach. In this approach, the width of the bin will vary over $\Gamma$ and also there is a constant robustness quality across all bins (Barra et al. 2014).

[7] Since our quantifier values are in the shape of fraction, we compute the averages by using the method of *global mediant*. In the method, the averages are obtained by computing the ratio between the statistical average of nominators and the statistical average of denominators (see Barra et al. 2014).

**Table 1** Binning and Goodness-of-fit statistics on social integration quantifiers, Spain 1998–2008.

| Period | Total Data | The number of bins | Population in each bin | $r_F$ | $R^2_F$ | $r_I$ | $R^2_I$ |
|---|---|---|---|---|---|---|---|
| **Mixed Marriages** | | | | | | | |
| 1 | 2421 | 8 | 303 | 1.242 | 0.6997 | 0.3402 | 0.9767 |
| 2 | 4885 | 11 | 444 | 1.329 | 0.8703 | 0.3725 | 0.9164 |
| 3 | 7432 | 11 | 676 | 1.21 | 0.8591 | 0.3773 | 0.9373 |
| 4 | 10039 | 13 | 772 | 1.077 | 0.8384 | 0.4007 | 0.9757 |
| 5 | 12676 | 15 | 845 | 1.201 | 0.8539 | 0.4223 | 0.9816 |
| 6 | 15334 | 18 | 852 | 1.162 | 0.7977 | 0.4502 | 0.9904 |
| 7 | 18060 | 19 | 951 | 1.156 | 0.8076 | 0.4738 | 0.9882 |
| 8 | 20835 | 23 | 906 | 1.228 | 0.8476 | 0.4865 | 0.9893 |
| 9 | 23649 | 27 | 854 | 1.215 | 0.8512 | 0.5009 | 0.993 |
| 10 | 26546 | 28 | 941 | 1.187 | 0.8583 | 0.5242 | 0.9931 |
| **Newborns** | | | | | | | |
| 1 | 2421 | 11 | 220 | 0.8164 | 0.9272 | 0.3099 | 0.9637 |
| 2 | 4885 | 14 | 349 | 0.8595 | 0.9522 | 0.319 | 0.9615 |
| 3 | 7432 | 18 | 411 | 0.8167 | 0.9469 | 0.3333 | 0.9616 |
| 4 | 10039 | 19 | 523 | 0.7761 | 0.8874 | 0.3027 | 0.9772 |
| 5 | 12676 | 21 | 582 | 0.8067 | 0.9008 | 0.3012 | 0.9778 |
| 6 | 15334 | 23 | 659 | 0.7239 | 0.7857 | 0.2823 | 0.9731 |
| 7 | 18060 | 27 | 667 | 0.6895 | 0.7746 | 0.2828 | 0.9794 |
| 8 | 20835 | 30 | 690 | 0.6791 | 0.7826 | 0.2812 | 0.98 |
| 9 | 23649 | 30 | 780 | 0.6702 | 0.7906 | 0.2809 | 0.9803 |
| 10 | 26546 | 27 | 982 | 0.6484 | 0.8114 | 0.2831 | 0.9883 |

**Table 2** Binning and Goodness-of-fit statistics on socio-economic integration quantifiers, Spain 2005–2010.

| Period | Total Data | The number of bins | Population in each bin | $r_F$ | $R^2_F$ | $r_I$ | $R^2_I$ |
|---|---|---|---|---|---|---|---|
| **Permanent Jobs** | | | | | | | |
| 1 | 581 | 5 | 116 | 1.507 | 0.9758 | 0.431 | 0.7189 |
| 2 | 1165 | 11 | 106 | 1.555 | 0.9849 | 0.4658 | 0.7498 |
| 3 | 1684 | 12 | 140 | 1.569 | 0.9929 | 0.4896 | 0.7601 |
| 4 | 2279 | 10 | 228 | 1.56 | 0.9899 | 0.4882 | 0.7437 |
| 5 | 2875 | 12 | 240 | 1.515 | 0.9926 | 0.5014 | 0.7645 |
| 6 | 3471 | 14 | 248 | 1.515 | 0.9935 | 0.4988 | 0.7745 |
| **Temporary Jobs** | | | | | | | |
| 1 | 581 | 9 | 65 | 1.8 | 0.9264 | 0.4755 | 0.6574 |
| 2 | 1165 | 12 | 97 | 1.818 | 0.9579 | 0.5376 | 0.6770 |
| 3 | 1684 | 13 | 130 | 1.881 | 0.9596 | 0.5788 | 0.6772 |
| 4 | 2279 | 11 | 207 | 1.887 | 0.9613 | 0.5895 | 0.6809 |
| 5 | 2875 | 13 | 221 | 1.861 | 0.9699 | 0.6059 | 0.6976 |
| 6 | 3471 | 15 | 231 | 1.843 | 0.9770 | 0.6098 | 0.7179 |

The growth law of the integration process can then be expressed over the specified periods mathematically by using curve fitting tools over the obtained bins. The curve fitting process reveal that the data set on socio–economic quantifiers can be well predicted by linear model $r_F\Gamma$ while the nonlinear model $r_I\sqrt{\Gamma}$ successfully projects the social data–set across the analyzed time sequences. Hence our results agree with those reported by Barra et al. (2014). Thereafter, for simplicity of notation, we use $r$ instead of both $r_F$ and $r_I$. These findings shows that the values of coefficient $r$ recorded over the indicated periods converge to a fixed number $\tilde{r}$ and determine a curve which holds the formula:

$$r(t) = \tilde{r}\left(1 \pm e^{-\frac{t}{K}}\right), t = 2, \dots, t_f \tag{5}$$

where $t_f$ is referring to the final period and the $\tilde{r}$ and $K$ are parameter estimates describing the growth law using information from previous time periods (see tables 3 and 4).

**Table 3** Forecasting measures for social integration quantifiers, Spain 1998–2008.

| Mixed Marriages | | | |
|---|---|---|---|
| Period | $r$ | $\tilde{r}$ | $K$ |
| 1 | 0.3402 (0.2973, 0.3831) | --------------------------- | --------------------------- |
| 2 | 0.3725 (0.3017, 0.4432) | 0.3759 | 0.4247 |
| 3 | 0.3773 (0.3214, 0.4333) | 0.377 (0.3658, 0.3881) | 0.4303 (0.3383, 0.5222) |
| 4 | 0.4007 (0.3644, 0.4369) | 0.3865 (0.3504, 0.4227) | 0.4824 (0.1587, 0.806) |
| 5 | 0.4223 (0.3927, 0.452) | 0.3976 (0.3618, 0.4334) | 0.5452 (0.2018, 0.8886) |
| 6 | 0.4502 (0.4296, 0.4708) | 0.4111 (0.3711, 0.451) | 0.6267 (0.2191, 1.034) |
| 7 | 0.4738 (0.4509, 0.4967) | 0.4254 (0.3829, 0.4679) | 0.7269 (0.263, 1.191) |
| 8 | 0.4865 (0.4647, 0.5084) | 0.438 (0.3961, 0.4799) | 0.8266 (0.3347, 1.318) |
| 9 | 0.5009 (0.4835, 0.5183) | 0.4497 (0.4086, 0.4908) | 0.9287 (0.4107, 1.447) |
| 10 | 0.5242 (0.5066, 0.5418) | 0.4631 (0.4211, 0.5051) | 1.064 (0.4949, 1.634) |
| Newborns | | | |
| Period | $r$ | $\tilde{r}$ | $K$ |
| 1 | 0.3099 (0.27, 0.3498) | --------------------------- | --------------------------- |
| 2 | 0.319 (0.2852, 0.3528) | 0.3145 | 0.2686 |
| 3 | 0.3333 (0.298, 0.3686) | 0.3207 (0.1708, 0.4706) | 0.07791 (−1844, 1844) |
| 4 | 0.3027 (0.2791, 0.3263) | 0.3162 (0.2815, 0.351) | 0.03235 |
| 5 | 0.3012 (0.2794, 0.323) | 0.3132 (0.2914, 0.335) | 0.02362 |
| 6 | 0.2823 (0.2611, 0.3036) | 0.3076 (0.2836, 0.3317) | 0.208 (−0.8099, 1.226) |
| 7 | 0.2828 (0.2659, 0.2997) | 0.3034 (0.2822, 0.3246) | 0.2693 (−0.2819, 0.8205) |
| 8 | 0.2812 (0.2657, 0.2967) | 0.3001 (0.2812, 0.3189) | 0.3059 (−0.1312, 0.7431) |
| 9 | 0.2809 (0.2655, 0.2963) | 0.2975 (0.2807, 0.3143) | 0.3332 (−0.04511, 0.7115) |
| 10 | 0.2831 (0.2704, 0.2958) | 0.2958 (0.2809, 0.3107) | 0.3518 (0.01316, 0.6905) |

**Table 4** Forecasting measures for socio-economic integration quantifiers, Spain 2005–2010.

| Permanent Jobs | | |
|---|---|---|
| **Period** | **$r$** | **$\tilde{r}$** | **$K$** |
| 1 | 1.507 (1.410, 2.472) | ------------------------ | --------- |
| 2 | 1.555 (1.469, 1.642) | 1.507 (1.410, 2.472) | 0.0000 |
| 3 | 1.569 (1.501, 1.637) | 1.555 (1.469, 1.642) | 0.0000 |
| 4 | 1.56 (1.49, 1.63) | 1.569 (1.501, 1.637) | 0.0000 |
| 5 | 1.515 (1.468, 1.563) | 1.56 (1.49, 1.63) | 0.0000 |
| 6 | 1.515 (1.476, 1.554) | 1.515 (1.468, 1.563) | 0.0000 |
| **Temporary Jobs** | | |
| **Period** | **$r$** | **$\tilde{r}$** | **$K$** |
| 1 | 1.8 (1.514, 2.086) | ------------------------ | --------- |
| 2 | 1.818 (1.629, 2.007) | 1.8 (1.514, 2.086) | 0.0000 |
| 3 | 1.881 (1.66, 2.102) | 1.818 (1.629, 2.007) | 0.0000 |
| 4 | 1.887 (1.703, 2.071) | 1.881 (1.66, 2.102) | 0.0000 |
| 5 | 1.861 (1.732, 1.99) | 1.887 (1.703, 2.071) | 0.0000 |
| 6 | 1.843 (1.735, 1.951) | 1.861 (1.732, 1.99) | 0.0000 |

Based on this piece of evidence, we suggest the parameter $\tilde{r}$ as a predictor to estimate the coefficient $r$ of subsequent period. Thereafter, we insert the predictor of $r$ into our models, and proceed by predicting the future outcomes. Our aim is to predict the quantifier values for new observations given their immigrant densities. It is worth noticing that in our approach, the value of each integration quantifier in the coming years also considers the new immigrants density $\Delta\gamma$ and thus the new value becomes $Q_i(\Gamma) + Q_i(\Delta\Gamma), i = 1,2,...,4$; where $Q$ represents the extracted growth low function for each quantifier and $\Delta\Gamma = \Delta\gamma(1 - 2\gamma)$.

For instance, regarding equation 5, assuming that the reported values of "$r$" through all these periods of time is a series of numbers, "$r$" will converge to the fixed numbers 0.5212 & 0.2821 for the mixed marriages and births with mixed parents quantifiers and numbers 1.515 & 1.861 for the quantifiers measuring the coefficient numbers of permanent and temporary jobs market respectively. Thus, if we use these "$r(t)$" values as the set of points that determines the curve for the four quantifiers, the following formulas apply:

$$r_{M_m}(t) = \tilde{r}_{M_m}\left(1 - e^{-\frac{t}{K_{M_m}}}\right) \quad \tilde{r}_{M_m} = 0.5212, K_{M_m} = 0.05221$$

$$r_{B_m}(t) = \tilde{r}_{B_m}\left(1 + e^{-\frac{t}{K_{B_m}}}\right) \quad \tilde{r}_{B_m} = 0.2821, K_{B_m} = 0.1058$$

$$r_{J_p}(t) = \tilde{r}_{J_p}\left(1 \pm e^{-\frac{t}{K_{J_p}}}\right) \quad \tilde{r}_{J_p} = 1.515, K_{J_p} = 0.0000$$

$$r_{J_t}(t) = \tilde{r}_{J_t}\left(1 \pm e^{-\frac{t}{K_{J_t}}}\right) \quad \tilde{r}_{J_t} = 1.861, K_{J_t} = 0.0000$$

where $t = 1, \ldots, t_f$. As a result, the values at convergence are introduced into our models for predicting future outcomes:

$$M_m(\Gamma) = 0.5212\sqrt{\Gamma} \qquad for\ Mixed\ marriages$$
$$B_m(\Gamma) = 0.2821\sqrt{\Gamma} \quad for\ Newborns\ with\ mixed\ parents$$

$$J_p(\Gamma) = 1.515\Gamma \qquad for\ Permanent\ contracts$$
$$J_t(\Gamma) = 1.861\Gamma \qquad for\ Temporary\ contracts$$

The quality of the subsequent forecasts turns out to be prompt, accurate and robust. Figure 1 illustrates how the goodness of fit of our forecasts changes as we progressively add information from subsequent time periods. As shown in the Figure 1, this curve fitting exercise tells us the growth law that provides the best fit, when the entire set of data is considered, visibly delivers substantially better forecasts as early on as from period one. Even in the worst case scenario (see upper right panel in the Figure 1) the forecast obtained by applying the square root growth law deliver more efficient estimates than the linear growth law already from period two and beyond. Thus, the predictability with respect to discerning the underlying mechanism for integration is prompt since the correct behavior emerges from the dataset as early as the first year, and no later than after two years.

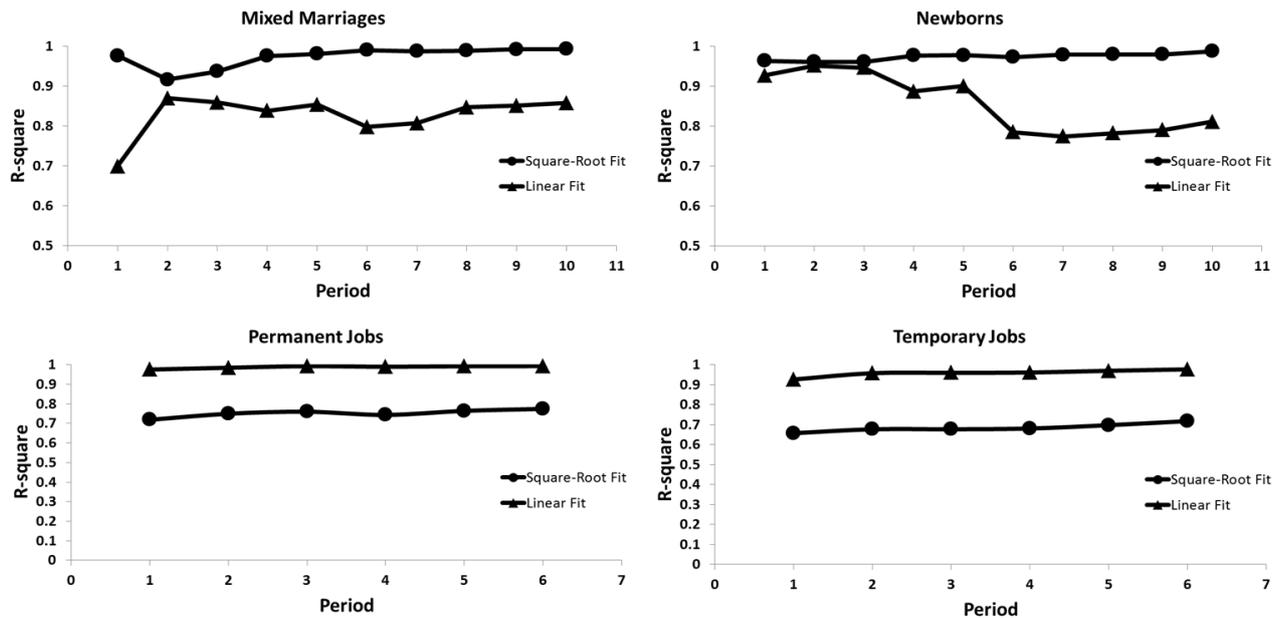

**Figure 1** Here each circular bullet displays the coefficient of determination computed for square–root fitting $r_I\sqrt{\Gamma}$ in the relevant period and each triangular bullet represents the analogous value recorded for linear model $r_F\Gamma$.

Figure 2 shows how the forecasted values $\tilde{r}$ (grey points) based on past data match the values $r$ observed in the successive year. The match turns out to be very accurate. The Adjusted Mean Absolute Percentage Errors[8] are within 1%.

---

[8] We used the Adjusted Mean Absolute Percentage Error which is a valid quantity to show the accuracy of the predictions:

adjusted MAPE $= \frac{1}{n}\sum_{t=1}^{n}\frac{|f_t - y_t|}{(f_t + y_t)/2}$ where $y_t$ and $f_t$ are observed and forecasting values respectively (see Armstrong 1985).

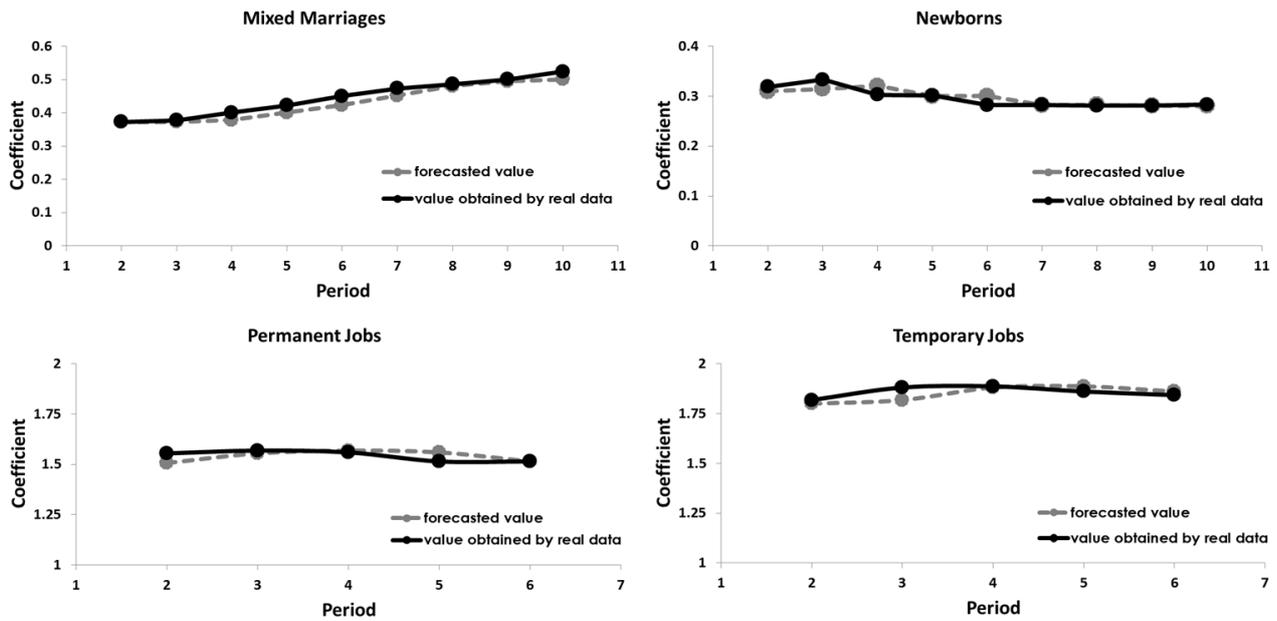

**Figure 2** The black points represent the value of the coefficient $r$ (for the suitable growth law identified in Figure 1) obtained by real data collected up to the indicated year, grey points represent the forecasted value of the same coefficient obtained with the equation $5$ – the Adjusted Mean Absolute Percentage Errors do are always under $1\%$.

Figure 3, finally, analyzes the robustness of the forecasting ability with respect to identified models over the entire time span. The fit of the forecasted value vis–a–vis the observed value remain at an exceptionally high level also for large time intervals. The integration quantifiers exhibiting linear growth behavior are identified with high accuracy, with an $R^2$ that never goes below .98 in the case of permanent contracts and .94 for temporary contracts. [9] As for the two quantifiers evolving like the square root, the lowest $R^2$ is .91 for Mixed marriages, whereas in the case of newborns with mixed parents it is .95.

---

[9] $R^2$ indicates how well the observed outcomes are replicated by the statistical model. The measure ranges from 0 to 1 such that the larger numbers representing better fits and also 1 indicates a perfect fit:

$$R^2 = 1 - \frac{\sum_{t=1}^{n}(y_t - f_t)^2}{(y_t - \bar{y})^2}$$

where $y_1, \dots, y_n$ are the observed values, $f_1, \dots, f_n$ are the forecasts, and $\bar{y}$ is the average of the data (Draper and Smith 1998).

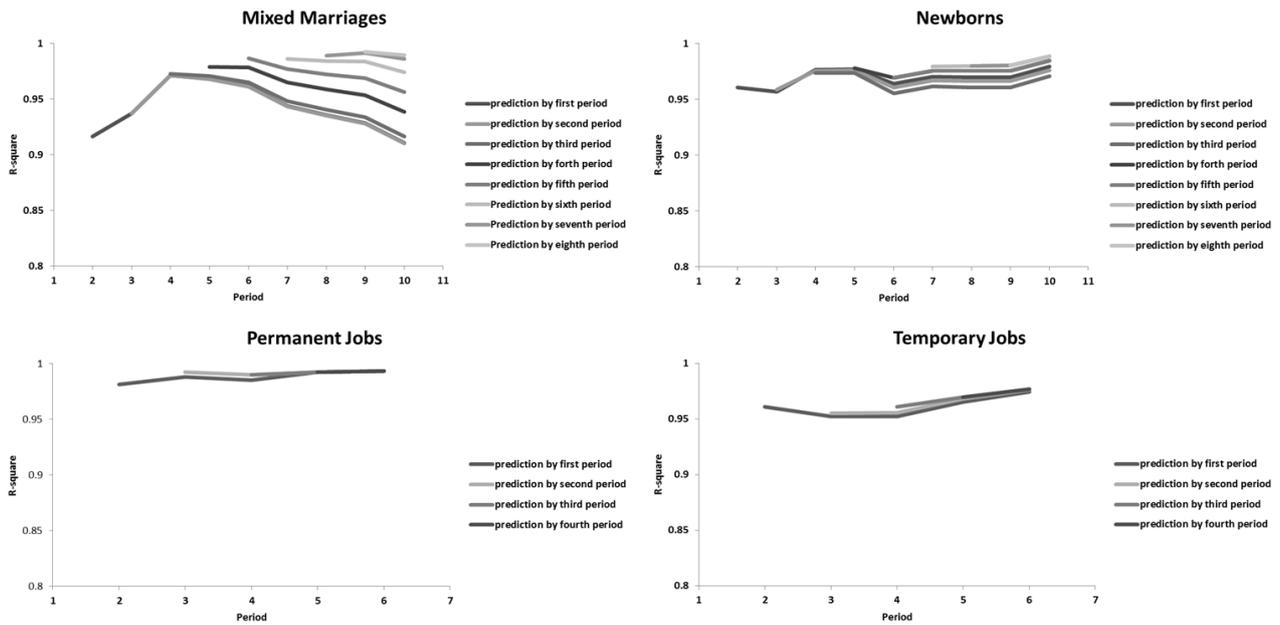

**Figure 3** Forecasts by the equation 5 are tested by the $R^2$ coefficient against real data at increasing time intervals. $R^2$ never goes below .91 even for forecasts 9 years ahead.

## Concluding discussion

The main objective of this work was to propose a quantitative method capable of elaborating precise forecasts of immigrant integration as a function of immigrant density. This objective has been accomplished.

The forecasting ability of the proposed model turns out to be prompt, accurate and robust. *Promptnes*s is found since the applied method successfully determines and differentiates between integration processes sensitive to social interaction or not, at very early stages (maximum two time periods). *Accuracy* and *Robustness* is found since, once we have determined the growth law, our prediction algorithm provide very accurate forecasts over the entire time sequence. Hence, using this framework it is possible to estimate the future rate of, for example, inter–marriage, if the size of the immigrant population rise by say 2–3%.

An important quality of our forecasting framework is that it is capable of uncovering the underlying mechanism driving the integration process—social interaction or independent decision making—at an early stage in the immigration cycle. The capacity to foretell information of this type is by no means trivial since the two mechanisms are likely to demand different policy responses. For example, when integration grow linearly with immigrant density, as our labor participation indicators, effective policy responses should focus on problems such as access to labor markets, to improve integration, whereas integration induced by social action and interaction requires policies targeting the quality and intensity of interaction between immigrants and locals.

Our findings are of particular value to governments and researchers engaged in formulating more effective and precise immigration and integration policies. It is also an excellent research tool for scholars interested in explaining individual integration outcomes since it unveils in a general but powerful way the presence or absence of social network effects in integration phenomena.

## References

Alba, R. and Nee, V. 1997. Rethinking assimilation theory for a new era of immigration. *International migration review*, 31(4):826–74.

Armstrong, J. S. 1985. *Long–range forecasting: from crystal ball to computer*. Wiley, 2nd edition.

Ballerini M, Cabibbo N, Candelier R, Cavagna A, Cisbani E, Giardina I, Lecomte V, Orlandi A, Parisi G, Procaccini A, Viale M and Zdravkovic V. (2008). Interaction ruling animal collective behavior depends on topological rather than metric distance: Evidence from a field study, *Proc. Natl. Aca. Sc.*, 105, 1232–1237.

Barra A. and Contucci, P. 2010. Toward a quantitative approach to migrants social integration. *EPL (Europhysics Letters)*, 89 (6), 68001.

Barra A., Contucci, P., Sandell, R. and Vernia, C. 2014. An analysis of a large dataset on immigrant integration in Spain. The statistical mechanics perspective on social action. *Scientific Reports* 4, 4174.

Bialek, W., Cavagna, A., Giardina, I., Mora, T., Silvestri, E., Viale, M. and Walczak A.M. 2012. Statistical mechanics for natural flocks of birds. *Proc. Natl. Acad. Sc.*, 109, 4786–4791.

Blau, P.M., Blum, T.C. and Schwartz, J.E. 1982. Heterogeneity and Inter–marriage. *American Sociological Review*. 47(1), 45–62.

Blau, P.M., Beeker, C. and Fitzpatrick, K.M. 1984. Intersecting social affiliations and inter–marriage. *Social Forces*, 62(3), 585–606.

Bouchaud, J.P. and Potters, M. 2004. *Theory of financial risks: from statistical physics to risk management*. Oxford University Press.

Brock, W. and Durlauf, S.N. 2001. Discrete choice with social interactions, *The Review of Economic Studies*, 68(2), 235–260.

Canada Government. 2012. Canada's economic action plan (2012), *a fast and flexible economic immigration system, jobs growth and long–term prosperity*. Technical report, Government of Canada.

Castellano, C., Marsili, M. and Vespignani, A. 2000. Nonequilibrium phase transition in a model for social influence. *Physicl Review Letters*, 85(6), 3536.

Castles, S. and Miller, M. J. (eds) 2009. *The age of migration – International population movements in the modern world*, Pallgrave McMillian, New York.

Contucci, P., Gallo, I. and Menconi, G. 2008. Phase transitions in social sciences: two–populations mean field theory, *International Journal of Modern Physics B*. 22(14). P 1–14.

Contucci P, and Giardina, C. 2008. Mathematics and Social Sciences: A statistical mechanics approach to immigration. *ERCIM News*. 73, 8241.

Cortina, C. T., Esteve, A. and Domingo, A. 2008. Marriage patterns of the foreign-born population in a new country of immigration: The case of Spain1. *International Migration Review*, 42(4), 877–902.

Draper, N. R. and Smith, H. 1998. *Applied Regression Analysis*. Wiley–Interscience.

European Commission. 2005. *A common agenda for integration framework for the integration of third–Country nationals in the European Union*. Technical Report COM (389), Commission of the European Communities,

European Commission. 2010. *EUROPE 2020: A strategy for smart, sustainable and inclusive growth*. COM (2020).

European Commission. 2011. *The global approach to migration and mobility*. EU Report, Technical Report COM(743).Sec(1353).

European Commission. 2014. *EU actions to make integration work: Common basic principles, European Website on Integration*, Technical report.

European Council. 2010. *The stockholm programme–an open and secure europe serving and protecting citizens official journal*, Technical Report C(115/1–38).

Gallo, I., Barra, A. and Contucci, P. 2009. Parameter evaluation of a simple mean–field model of social interaction. *Math. Models Methods Appl. Sci*. 19, 1427.

Giovagnoli, M. 2011. *Improving the naturalization process: Better immigrant integration lead to economic growth*. Immigration Policy Center, American Immigration Council.

Gordon, M.M. 1964. *Asimilation in American Life: The role of Race, Religion, and National Origins*. Oxford University Press. New York

Granovetter, M. 1974. *Getting a Job: A study of contacts and careers*. The University of Chicago Press. Chicago.

Heilmann, O.J. and Lieb, E.H. 1970. Monomers and dimers, *Physical Review Letters*, 24, 1412–1414.

Hwang, S.S., Saenz, R. and Aguirre, B.E. (1997). Structural and assimilationist explanations of Asian American inter–marriage. *Journal of Marriage and the Family*, 59(3), 758–72.

IOM. 2011. *International dialogue on migration N. 17 – Migration and social change*. The International Dialogue on Migration Red Book Series, Geneva, Switzerland.

Jacoby T. Culver C., Daley R.M., and Munana C. 2013. *US economic competitiveness at risk: A midwest call to action on immigration reform*. The Chicago Council on Global Affairs, Chicago, Illinois.

Kalmijn, M. 1998. Inter–marriage and Homogamy: Causes, Patterns, Trends. *Annual review of sociology* 24(24), 395–421.

Levy M 2005. Social phase transition. *Journal of Economic Behavior & Organization*, 57. P 71–87.

Mantegna, R.N. and Stanley, H.E. 1999. *Introduction to econophysics: correlations and complexity in finance*. Cambridge University Press.

McFadden, D. 2001. Economic choices, *The Amer. Econ. Rev.*, 91, 351–378.

Niessen J. and Huddleston T. (eds) 2010. *Handbook on integration for policy–makers and practitioners*. Migration Policy Group on behalf of the European Commission, Farance.

Penninx R., Spencer, D. and Hear, N.V. 2008. *Migration and integration in Europe: The state of research*. ESRC Centre on Migration, Policy and Society (COMPAS) University of Oxford.

Portes, A. and M. Zhou. 1993. The new second generation: Segmented assimilation and its variants. *The Annals of the American Academy of Political and Social Science*, 530.

Qian, Z., Glick, J.E., and Batson, C.D. 2012. Crossing boundaries: Nativity, Ethnicity, and Mate Selection. *Demography*, 49(2), 651–75.

Raijman R., and Tienda. M. 1999. Immigrants socio–economic progress post–1965: forgingmobility or survival? In *The Handbook of International Migration*. Russell Sage Foundation, New York.

Sandell, R. 2012. Social influences and aggregated immigration dynamics: The case of Spain 1999–2009, *International Migration Review, Wiley Blackwell*, 46(4), 971–1004.

Stanley. H.E. 2008. Econophysics and the current economic turmoil. *American Physical Society News*, 11(17):8.

Stevens, G. and Swicegood, G. 1987. The linguistic context of ethnic endogamy." *American Sociological Review*, 52(1):73–82. Retrieved from http://www.jstor.org/stable/2095393

Van Tubergen. F. 2006. *Immigrant integration: A cross–national study*. LFB Scholarly Publishing LLC.

Van Tubergen F. and Maas. I. 2007. Ethnic inter–marriage among immigrants in the Netherlands: An analysis of population data. *Social Science Research*, 36(3):106586.

Watts, D. J. and Strogatz, S. H. 1998. Collective dynamics of small–world networks. *Nature*, 393 (6684), 440–442.

Weber, M. 1978. *Economy and Society: An outline of interpretive sociology*. University of California Press., 23.